\documentclass[conference]{IEEEtran}

\ifCLASSINFOpdf

\else

\fi

\usepackage{algpseudocode}
\usepackage{algorithm}
\usepackage[dvips]{graphicx}
\usepackage{latexsym}
\usepackage{amssymb}
\usepackage{amsmath}
\usepackage{multirow}
\usepackage{xcolor}
\usepackage{lipsum}
\usepackage{multicol}
\usepackage{enumerate}
\usepackage{algorithm}
\usepackage{algorithmicx}
\usepackage{algpseudocode}
\usepackage{amsmath}
\usepackage{graphicx}
\usepackage{subfig}
\usepackage[noadjust]{cite}

\begin{document}

\title{1-Bit Massive MIMO Downlink Based on Constructive Interference}

\author{\IEEEauthorblockN{Ang Li$^{\dagger}$, Christos Masouros$^{\dagger}$, and A. Lee Swindlehurst$^*$}
\IEEEauthorblockA{Dept. of Electronic and Electrical Eng., University College London, London, U.K.$^{\dagger}$\\
Center for Pervasive Communications and Computing, University of California Irvine, Irvine, CA 92697, USA$^*$\\
Email: \{ang.li.14, c.masouros\}@ucl.ac.uk$^{\dagger}$, swindle@uci.edu$^*$}
}

\maketitle

\begin{abstract}
In this paper, we focus on the multiuser massive multiple-input single-output (MISO) downlink with low-cost 1-bit digital-to-analog converters (DACs) for PSK modulation, and propose a low-complexity refinement process that is applicable to any existing 1-bit precoding approaches based on the constructive interference (CI) formulation. With the decomposition of the signals along the detection thresholds, we first formulate a simple symbol-scaling method as the performance metric. The low-complexity refinement approach is subsequently introduced, where we aim to improve the introduced symbol-scaling performance metric by modifying the transmit signal on one antenna at a time. Numerical results validate the effectiveness of the proposed refinement method on existing approaches for massive MIMO with 1-bit DACs, and the performance improvements are most significant for the low-complexity quantized zero-forcing (ZF) method.
\end{abstract}


\IEEEpeerreviewmaketitle

\section{Introduction}
Massive multiple-input multiple-output (MIMO) systems have been recognized as a promising technology for the fifth generation (5G) and future wireless communications \cite{r1}. With channel state information (CSI) at the base station (BS), fully-digital massive MIMO systems can greatly improve the spectral efficiency by employing simple linear precoding schemes, where it has been shown that zero-forcing (ZF)-based precoding approaches are near-optimal when favorable propagation conditions exist \cite{r1}. Nevertheless, with a fully-digital large-scale antenna array at the BS, the corresponding large number of radio frequency (RF) chains and high-resolution digital-to-analog converters (DACs) lead to greatly increased hardware complexity, cost, and power consumption at the BS, which hinders the implementation of massive MIMO in practical wireless systems. One popular technique that achieves a performance-cost compromise is to reduce the number of RF chains by employing hybrid structures \cite{r2}\nocite{r3}\nocite{r17}\nocite{r18}-\cite{r25}, where the precoding is divided into the analog domain and digital domain. With a reduced number of RF chains, the hardware costs and power consumption at the BS can be reduced accordingly.

In addition to the hybrid structures, another promising technique, which is the focus of this paper, is to reduce the power consumption per RF chain by employing low-resolution DACs. It is known that the power consumption of DACs grows exponentially with the resolution and linearly with the bandwidth \cite{r4}. Therefore, employing low-resolution DACs, especially 1-bit DACs, can greatly alleviate the hardware cost and the power consumption at the BS in the downlink. Due to the benefits that 1-bit DACs can provide, in the recent literature there are an increasing number of  contributions that propose 1-bit transmit precoding designs for the massive MIMO downlink \cite{r5}\nocite{r6}\nocite{r7}\nocite{r8}\nocite{r9}\nocite{r10}\nocite{r11}\nocite{r12}\nocite{r13}-\cite{r23}. In \cite{r5}, the 1-bit quantization is directly applied to the linear ZF precoding and its performance is analytically studied. In \cite{r6}\cite{r7}, linear precoding schemes based on the minimum-mean-squared error (MMSE) criterion are proposed. However, the above quantized linear precoding methods are shown to suffer severe performance losses in the medium-to-high signal-to-noise ratio (SNR) regime due to the 1-bit quantization. 

To further improve the system performance in the presence of 1-bit DACs, non-linear precoding approaches that directly design the transmit signals based on the data symbols are further proposed in \cite{r8}-\cite{r13}, where a symbol perturbation scheme for QPSK modulation is  introduced in \cite{r8}, \cite{r9} proposes a non-linear precoding method based on the gradient descend method (GDM), while other non-linear precoding methods are proposed in \cite{r10}\cite{r11} based on biconvex relaxation. More computationally intensive non-linear methods have been proposed in \cite{r12}-\cite{r23} that achieve better performance, but with a complexity that scales at least as the square of the number of antennas. While such approaches offer better performance compared with the low-complexity quantized linear precoding methods, their complexity may be too high for practical implementation.

To improve the performance of the more practical low-complexity quantized linear precoding methods, we propose a low-complexity refinement method based on the constructive interference (CI) formulation, which is applicable to any existing 1-bit precoders for PSK modulation. By decomposing the data symbols and received signals along the corresponding detection thresholds, and using an appropriate coordinate transformation, we first construct a simple symbol-scaling formulation as our performance metric based on CI. Based on this metric, we further introduce a refinement that improves the symbol-scaling metric of existing quantized precoding schemes by modifying the transmit signal one antenna at a time. Our numerical results validate the effectiveness of the proposed refinement method on a number of existing quantized precoding approaches, and reveal the fact that the performance improvements are more significant for the quantized linear methods.

$Notations$: $a$, $\bf a$, and $\bf A$ denote scalar, vector and matrix, respectively. ${( \cdot )^T}$, ${( \cdot )^H}$, and $tr\left\{  \cdot  \right\}$ denote transposition, conjugate transposition, and trace of a matrix respectively. $\left|  \cdot  \right|$ denotes the modulus of a complex number, and ${{\cal C}^{n \times n}}$ represents the set of $n \times n$ complex matrices. $\Re ( \cdot )$ and $\Im ( \cdot )$ denote the real and imaginary part of a complex number, respectively.

\begin{figure}[h]
\centering
\includegraphics[scale=0.245]{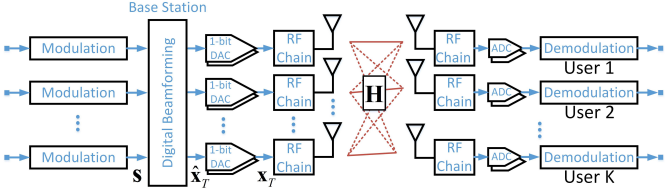}
\caption{A massive MIMO downlink system with 1-bit DACs}
\end{figure}

\section{System Model}
We consider a multiuser massive MISO downlink with 1-bit DACs employed at the BS, where the BS with $N_t$ transmit antennas communicates with $K$ single-antenna users simultaneously in the same time-frequency resource and $K \le N_t$, as depicted in Fig. 1. We focus on transmit precoding designs with 1-bit DACs, and ideal analog-to-digital converters (ADCs) with infinite precision are assumed for each receiver. Following the closely-related literature \cite{r5}-\cite{r9}, the data symbols are assumed to be from a normalized PSK modulation, denoted as ${\bf s}\in {\cal C}^{K\times1}$. We denote the quantized transmit signal vector as ${\bf x}_T \in {\cal C}^{N_t\times1}$, and ${\bf x}_T$ can be expressed as
\begin{equation}
{{\bf{x}}_T} = {\cal Q}\left\{ {{\cal B}\left( {\bf{s}} \right)} \right\}.
\end{equation}
In (1), $\cal B$ denotes a generic precoding strategy, where for a linear precoding scheme $\cal B$ is equivalent to multiplication by a precoding matrix, while for non-linear precoding approaches $\cal B$ denotes the mapping strategy to form the unquantized signals based on $\bf s$. $\cal Q$ denotes the element-wise 1-bit quantization operation for both the real and imaginary part, and in this paper we normalize each entry in ${\bf x}_T$ as
\begin{equation}
x_n \in \left\{ { \pm \frac{1}{{\sqrt {2N_t} }} \pm j \frac{1}{{\sqrt {2N_t} }}} \right\}, \forall n \in {\cal N},
\end{equation}
where ${\cal N} = \left\{ {1,2, \cdots ,{N_t}} \right\}$. The above normalization guarantees that ${\left\| {{{\bf{x}}_T}} \right\|_F} = 1$. Subsequently, we express the received signal for the $k$-th user as
\begin{equation}
\begin{aligned}
y_k&=\sqrt P \cdot {\bf h}_k{{\bf{x}}_T}+n_k \\
&=\sqrt P \cdot {\bf h}_k{\cal Q}\left\{ {{\cal B}\left( {\bf{s}} \right)} \right\}+n_k,
\end{aligned}
\end{equation}
where ${\bf h}_k$ denotes the flat-fading Rayleigh fading channel vector for user $k$ with each entry following a standard Gaussian distribution, $n_k$ is the additive Gaussian noise at the receiver with zero mean and variance $\sigma^2$, and $P$ denotes the total available power per antenna, where for simplicity we have assumed a uniform power allocation.

\begin{figure}[!b]
\centering
\includegraphics[scale=0.35]{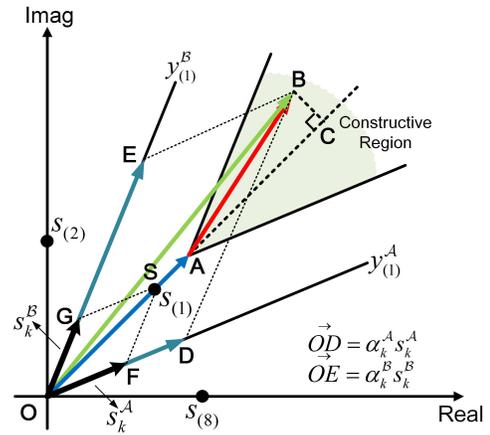}
\caption{8PSK as an example for the signal decomposition}
\end{figure}

\section{Proposed Refinement Method based on CI}
In this section, the symbol-level performance metric based on constructive interference is first introduced, followed by the proposed refinement process.

\subsection{Constructive Interference}
We define CI as interference that pushes the received symbols away from the detection thresholds of the modulation constellation \cite{r14}\nocite{r15}-\cite{r16}. The exploitation of CI was first introduced in \cite{r14} and further extended in \cite{r16} with the introduction of the constructive region. These papers showed that, as long as the resulting interfered signals are located in the constructive region, the distance to the decision thresholds is increased, and the introduction of the constructive region brings additional performance gains \cite{r16}. While we focus on PSK modulation in this paper, the extension to QAM modulations are applicable, and we refer the interested readers to \cite{r21}-\cite{r24} for a detailed description.

\subsection{The Symbol-Scaling Performance Metric}
Based on the above description for CI, we propose to consider an alternative symbol-scaling metric for the 1-bit quantized massive MIMO downlink based on the CI formulation, which characterizes the effect of the interference on each data symbol in a simple way. To be specific, for a generic $\cal M$-PSK modulation, we first express each data symbol as 
\begin{equation}
{s_{\left( l \right)}} = {e^{j \cdot \left[ {\frac{{2\pi }}{\cal M}\left( {l - 1} \right) + \frac{\pi }{4}} \right]}}, {\kern 3pt} l \in \left\{ {1,2, \cdots {\cal M}} \right\},
\end{equation}
where ${s_{\left( l \right)}}$ denotes the $l$-th constellation point which starts from ${e^{j \cdot \frac{\pi }{4}}}$ and follows an anti-clockwise direction, as depicted in Fig. 2 where we employ 8PSK as an example. We then decompose each data symbol $s_k$ along the corresponding detection thresholds, where without loss of generality we assume ${s_k} = {s_{\left( l \right)}}$, which leads to
\begin{equation}
{s_k} = {s_{\left( l \right)}} = s_k^{\cal A} + s_k^{\cal B}.
\end{equation}
Based on the geometry, we can obtain the following expression for the decomposed signal $s_k^{\cal A}$ and $s_k^{\cal B}$
\begin{equation}
\begin{aligned}
s_k^{\cal A} &= \frac{{{e^{j \cdot \left[ {\frac{{2\pi }}{\cal M} \cdot \left( {l-1} \right) +\frac{\pi}{4} - \frac{{\pi }}{\cal M}} \right]}}}}{\rho } = A_k^\Re  + j \cdot A_k^\Im, \\
s_k^{\cal B} &= \frac{{{e^{j \cdot \left[ {\frac{{2\pi }}{\cal M} \cdot \left( {l-1} \right) +\frac{\pi}{4} + \frac{\pi }{\cal M}} \right]}}}}{\rho } = B_k^\Re  + j \cdot B_k^\Im ,
\end{aligned}
\end{equation}
where $\left( {A_k^\Re ,A_k^\Im } \right)$ and $\left( {B_k^\Re ,B_k^\Im } \right)$ denote the coordinates of the bases $s_k^{\cal A}$ and $s_k^{\cal B}$ in the real-imaginary plane, respectively, and $\rho=$ is a scaling factor to ensure the equality of (5):
\begin{equation}
\rho  = \left| {{e^{j \cdot \left[ {\frac{{2\pi }}{\cal M} \cdot \left( {l-1} \right) +\frac{\pi}{4} - \frac{{\pi }}{\cal M}} \right]}} + {e^{j \cdot \left[ {\frac{{2\pi }}{\cal M} \cdot \left( {l-1} \right) +\frac{\pi}{4} + \frac{\pi }{\cal M}} \right]}}} \right|.
\end{equation}
For example, for the constellation point ${s_{\left( 1 \right)}}$ considered in Fig. 2, we obtain the expressions
\begin{equation}
\begin{aligned}
s_k^{\cal A} &= \frac{{{e^{j \cdot \left( {\frac{{2\pi }}{8} \cdot 0 +\frac{\pi}{4} - \frac{{\pi }}{8}} \right)}}}}{\left| {{e^{j \cdot \left( {\frac{{2\pi }}{8} \cdot 0 + \frac{\pi }{4} - \frac{\pi }{8}} \right)}} + {e^{j \cdot \left( {\frac{{2\pi }}{8} \cdot 0 + \frac{\pi }{4} + \frac{\pi }{8}} \right)}}} \right|} = \frac{{{e^{j \cdot \frac{\pi }{8}}}}}{{\left| {{e^{j \cdot \frac{\pi }{8}}} + {e^{j \cdot \frac{{3\pi }}{8}}}} \right|}}, \\
s_k^{\cal B} &= \frac{{{e^{j \cdot \left( {\frac{{2\pi }}{8} \cdot 0 +\frac{\pi}{4} + \frac{\pi }{8}} \right)}}}}{\left| {{e^{j \cdot \left( {\frac{{2\pi }}{8} \cdot 0 + \frac{\pi }{4} - \frac{\pi }{8}} \right)}} + {e^{j \cdot \left( {\frac{{2\pi }}{8} \cdot 0 + \frac{\pi }{4} + \frac{\pi }{8}} \right)}}} \right|} = \frac{{{e^{j \cdot \frac{{3\pi }}{8}}}}}{{\left| {{e^{j \cdot \frac{\pi }{8}}} + {e^{j \cdot \frac{{3\pi }}{8}}}} \right|}}.
\end{aligned}
\end{equation}

Subsequently, we decompose the noiseless received signal of each user $k$ along the two detection thresholds of $s_k$:
\begin{equation}
{{\bf{h}}_k}{{\bf{x}}_T} = \alpha _k^{\cal A} s_k^{\cal A}  + \alpha _k^{\cal B} s_k^{\cal B},
\end{equation}
where 
\begin{equation}
\alpha _k^{\cal A} \ge 0, {\kern 3pt} \alpha _k^{\cal B} \ge 0, {\kern 3pt} \forall k \in \left\{ {1,2,\cdots, K} \right\},
\end{equation}
are real scaling factors for $s_k^{\cal A}$ and $s_k^{\cal B}$, respectively. A larger value for $\alpha _k^{\cal A}$ or $\alpha _k^{\cal B}$ represents a larger distance to the detection threshold, and accordingly the performance of 1-bit quantized MIMO systems is dominated by the minimum value of $\alpha _k^{\cal A}$ and $\alpha _k^{\cal B}$. Based on this fact, we employ the minimum value in the real scaling factors $\alpha _k^{\cal A}$ and $\alpha _k^{\cal B}$ as the performance metric.

\subsection{Proposed Refinement Method}
Before we introduce the refinement method, based on (6) and (9), we first obtain $\alpha _k^{\cal A}$ and $\alpha _k^{\cal B}$ as a function of the quantized transmit signal vector ${\bf x}_T$. We expand (9) by its real and imaginary part, expressed as
\begin{equation}
\begin{aligned}
{{\bf{h}}_k}{{\bf{x}}_T} &= \Re \left( {{{\bf{h}}_k}{{\bf{x}}_T}} \right) + j \cdot \Im \left( {{{\bf{h}}_k}{{\bf{x}}_T}} \right) \\
&=\left( {{\bf{h}}_k^\Re {\bf{x}}_T^\Re  - {\bf{h}}_k^\Im {\bf{x}}_T^\Im } \right) + j \cdot \left( {{\bf{h}}_k^\Im {\bf{x}}_T^\Re  + {\bf{h}}_k^\Re {\bf{x}}_T^\Im } \right) \\
&=\alpha _k^{\cal A}\left( {A_k^\Re  + j \cdot A_k^\Im } \right) + \alpha _k^{\cal B}\left( {B_k^\Re  + j \cdot B_k^\Im } \right),
\end{aligned}
\end{equation}
where for simplicity we have employed the following notations 
\begin{equation}
{\bf{x}}_T^\Re  = \Re \left( {{{\bf{x}}_T}} \right), {\kern 2pt} {\bf{x}}_T^\Im  = \Im \left( {{{\bf{x}}_T}} \right), {\kern 2pt} {\bf{h}}_k^\Re  = \Re \left( {{{\bf{h}}_k}} \right), {\kern 2pt} {\bf{h}}_k^\Im  = \Im \left( {{{\bf{h}}_k}} \right).
\end{equation} 
Accordingly, we obtain \cite{r20}
\begin{equation}
\begin{aligned}
\alpha _k^{\cal A} &= \frac{{{B_k^\Im} {{{\bf{h}}_k^\Re}} - {B_k^\Re}  {{{\bf{h}}_k^\Im}} }}{{{A_k^\Re}{B_k^\Im} - {A_k^\Im}{B_k^\Re}}}{\bf{x}}_T^\Re  - \frac{{{B_k^\Im} {{{\bf{h}}_k^\Im}} + {B_k^\Re} {{{\bf{h}}_k^\Re}} }}{{{A_k^\Re}{B_k^\Im} - {A_k^\Im}{B_k^\Re}}}{\bf{x}}_T^\Im, \\
\alpha _k^{\cal B} &=\frac{{{A_k^\Re}{{{\bf{h}}_k^\Im}} - {A_k^\Im} {{{\bf{h}}_k^\Re}}}}{{{A_k^\Re}{B_k^\Im} - {A_k^\Im}{B_k^\Re}}}{\bf{x}}_T^\Re  + \frac{{{A_k^\Re}{{{\bf{h}}_k^\Re}} + {A_k^\Im}{{{\bf{h}}_k^\Im}}}}{{{A_k^\Re}{B_k^\Im} - {A_k^\Im}{B_k^\Re}}}{\bf{x}}_T^\Im.
\end{aligned}
\end{equation}
By introducing 
\begin{equation}
\begin{aligned}
&{{\bf{A}}_k} = \frac{{{B_k^\Im}{{{\bf{h}}_k^\Re}} - {B_k^\Re} {{{\bf{h}}_k^\Im}} }}{{{A_k^\Re}{B_k^\Im} - {A_k^\Im}{B_k^\Re}}}, {\kern 5pt} {{\bf{B}}_k} =  - \frac{{{B_k^\Im} {{{\bf{h}}_k^\Im}} + {B_k^\Re}{{{\bf{h}}_k^\Re}} }}{{{A_k^\Re}{B_k^\Im} - {A_k^\Im}{B_k^\Re}}}, \\
&{{\bf{C}}_k} = \frac{{{A_k^\Re}{{{\bf{h}}_k^\Im}} - {A_k^\Im} {{{\bf{h}}_k^\Re}} }}{{{A_k^\Re}{B_k^\Im} - {A_k^\Im}{B_k^\Re}}}, {\kern 5pt} {{\bf{D}}_k} = \frac{{{A_k^\Re}{{{\bf{h}}_k^\Re}} + {A_k^\Im} {{{\bf{h}}_k^\Im}} }}{{{A_k^\Re}{B_k^\Im} - {A_k^\Im}{B_k^\Re}}},
\end{aligned}
\end{equation}
we simplify (13) into
\begin{equation}
\alpha _k^{\cal A} = {{\bf{A}}_k}{\bf{x}}_T^\Re  + {{\bf{B}}_k}{\bf{x}}_T^\Im, {\kern 5pt}
\alpha _k^{\cal B} = {{\bf{C}}_k}{\bf{x}}_T^\Re  + {{\bf{D}}_k}{\bf{x}}_T^\Im.
\end{equation}
By further introducing 
\begin{equation}
{{\bf{R}}_k} = \left[ {\begin{array}{*{20}{c}}
{{{\bf{A}}_k}}&{{{\bf{B}}_k}}
\end{array}} \right], {\kern 3pt} {{\bf{I}}_k} = \left[ {\begin{array}{*{20}{c}}
{{{\bf{C}}_k}}&{{{\bf{D}}_k}}
\end{array}} \right],
\end{equation}
and 
\begin{equation}
\begin{aligned}
&{\bf{x}}_E = {\left[ {\begin{array}{*{20}{c}}
{{{\left( {{\bf{x}}_T^\Re } \right)}^T}}&{{{\left( {{\bf{x}}_T^\Im } \right)}^T}}
\end{array}} \right]^T}, \\
&{\bf \Lambda}  = {\left[ {\alpha _1^{\cal A} ,\cdots,\alpha _K^{\cal A} ,\alpha _1^{\cal B} ,\cdots,\alpha _K^{\cal B} } \right]^T},
\end{aligned}
\end{equation}
(15) can be further expressed in a compact form as
\begin{equation}
{\bf \Lambda}  = {\bf{M}}{{\bf{x}}}_E,
\end{equation}
where $\bf M$ is given by
\begin{equation}
{\bf{M}} = {\left[ {\begin{array}{*{20}{c}}
{{{\bf{R}}_1^T}}& \cdots &{{{\bf{R}}_K^T}}&{{{\bf{I}}_1^T}}& \cdots &{{{\bf{I}}_K^T}}
\end{array}} \right]^T}.
\end{equation}
In (18), we denote $\alpha_l$ as the $l$-th entry in $\bf \Lambda$ and we omit the notations ${\cal A}$ and ${\cal B}$. Accordingly, the symbol-scaling performance metric is the minimum value of $\alpha_l$ in $\bf \Lambda$, which we aim to improve via the proposed refinement process.

\begin{algorithm}[b]
  \caption{The Proposed Refinement Method}
  \begin{algorithmic}
    \State ${\bf input:}$ ${\bf x}_T^0$ in existing 1-bit quantized precoding methods;
    \State ${\bf output:}$ ${\bf x}_T$;
    \State Calculate ${\bf{x}}_E^0 = {\left[ {\begin{array}{*{20}{c}}
{\Re {{\left( {{\bf{x}}_T^0} \right)}^T}}&{\Im {{\left( {{\bf{x}}_T^0} \right)}^T}}
\end{array}} \right]^T}$ via (17);
    \State Calculate $\bf M$ via (5)-(19);
    \For {$i = 1:2{N_t}$}
    \State Calculate ${\bf \Lambda}_0  = {\bf{M}}{{\bf{x}}_E^0}$;
    \State Obtain ${{\bf{x}}_E^i} = {\left[ {{x_E^1},\cdots,{x_E^{i-1}}, - {x_E^i},{x_E^{i+1}},\cdots,{x_E^{2N_t}}} \right]^T}$;
    \State Calculate ${\bf \Lambda}_i  = {\bf{M}}{{\bf{x}}_E^i}$;
    \If {$\min \left( {{{\bf \Lambda} _i}} \right) > \min \left( {{{\bf \Lambda} _0}} \right)$}
    \State ${x_E^i} \gets  - {x_E^i}$;
    \State Update ${\bf x}_E^0$;
    \EndIf
    \EndFor
    \State Obtain ${{\bf{x}}_T} = \left[ {\begin{array}{*{20}{c}}
{\bf{I}}&{j \cdot {\bf{I}}}
\end{array}} \right] \cdot {\bf{x}}_E^0$.
   \end{algorithmic}
\end{algorithm}

In the following we introduce the proposed refinement method based on the above formulation, which can be applied to any existing precoding methods for the 1-bit massive MIMO downlink to further improve performance. The central idea of the refinement scheme aims to improve the minimum value of $\alpha_l$ by modifying the transmit signals on the antennas. If we consider all the possible signal combinations, the refinement is equivalent to an exhaustive search, which is too computationally costly to implement. Therefore, to consider a more practical approach and keep the computational cost as low as possible, we propose to only modify the transmit signal on one antenna at a time. To be specific, we first calculate the scaling vector ${\bf \Lambda}_0$ based on ${\bf{x}}_E^0$ obtained from some existing algorithm via (5)-(18), and further obtain the minimum value of $\alpha_l$ in ${\bf \Lambda}_0$, which is denoted as $\min \left\{ {{\bf \Lambda}_0} \right\}$. Subsequently, we perform an iterative method, where within the $i$-th iteration, we change the sign of an entry in ${\bf x}_E$, and calculate the updated scaling vector ${\bf \Lambda}_i$. If the minimum value of $\alpha_l$ in the updated scaling vector, which is denoted as $\min \left\{ {{\bf \Lambda}_i} \right\}$, is larger than $\min \left\{ {{\bf \Lambda}_0} \right\}$, we update ${\bf x}_E$ accordingly. In the case that the minimum value of $\alpha_l$ becomes smaller, we keep the $i$-th entry in ${\bf x}_E$ unaltered and move to the next iteration. While the above approach is not guaranteed to converge to the global optimum, it will be shown that the performance improvement is indeed significant, especially for the quantized linear ZF methods.

Based on the above description, we summarize the refinement algorithm in Algorithm 1.

\section{Numerical Results}
To evaluate the performance of the proposed symbol scaling approach, in this section we present numerical results in terms of the bit error rate (BER) based on Monte Carlo simulations, where in each plot the transmit SNR is defined as $\rho={P \mathord{\left/ {\vphantom {P {{\sigma ^2}}}} \right.\kern-\nulldelimiterspace} {{\sigma ^2}}}$. Both QPSK and 8PSK modulations are considered in the simulations. The refinement method is simulated together with the following existing 1-bit quantized precoding methods: 

\begin{enumerate}

\item `ZF-Unquantized': the conventional ZF precoding with infinite-precision DACs (only as reference);
\item `ZF 1-Bit (R)': the 1-bit quantized ZF approach introduced in \cite{r5}\cite{r6};
\item `SP (R)': the 1-bit symbol perturbation precoding technique in \cite{r8} for QPSK;
\item `GDM (R)': GDM-based 1-bit precoding method in \cite{r9};
\item `C1PO (R)': the `C1PO' 1-bit precoding method in \cite{r11} based on biconvex-relaxation (equivalent to the `Pokemon' method in \cite{r10}) implemented with $n_{\max}=20$;
\item `Constructive (R)': the 1-bit precoding method based on CI in \cite{r20}.

\end{enumerate}
The inclusion of `R' in the above abbreviations denotes the cases where the proposed refinement method is applied to the original 1-bit precoding techniques.

\begin{figure}[!t]
\centering
\includegraphics[scale=0.45]{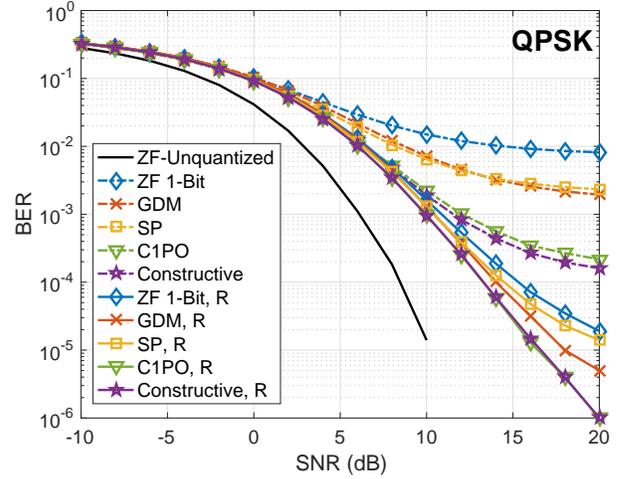}
\caption{BER v.s. transmit SNR, $N_t=8$, $K=2$, QPSK}
\end{figure}

In Fig. 3, we present numerical results for a small-scale MIMO system with $N_t=8$ transmit antennas and $K=2$ users. When the proposed refinement method is not applied, we observe that quantized non-linear precoding approaches generally offer a better BER performance than the quantized linear methods, and the quantized linear ZF scheme has the worst BER performance. Specifically, an error floor is observed for all existing precoding approaches in the high SNR regime, which is due to the 1-bit quantization. When the refinement process is applied to the existing algorithms, we observe significant improvements in the BER performance for all techniques, and the error floor vanishes for most of the approaches. Specifically, the proposed refinement process offers the highest performance improvement for the low-complexity quantized linear ZF method.

\begin{figure}[!t]
\centering
\includegraphics[scale=0.45]{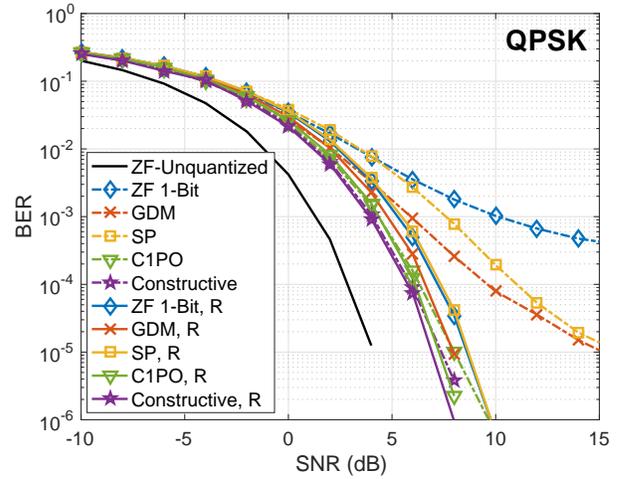}
\caption{BER v.s. transmit SNR, $N_t=128$, $K=16$, QPSK}
\end{figure}

In the following, we show the numerical results for massive MIMO systems. In Fig. 4, we present the BER performance for QPSK modulation with $N_t=128$ transmit antennas and $K=16$ users. An improved BER performance is observed compared to the case of small-scale MIMO systems in Fig. 3, due to the increase in the ratio $N_t/K$. Again, when the refinement method is not introduced, the low-complexity quantized linear ZF scheme achieves the worst BER performance and an error floor is observed. The more complicated non-linear methods achieve significantly better BER performance. When the refinement process is further introduced, the BER performance of all techniques is further improved, and the scheme with the worst BER performance improves the most. Specifically, we observe that, with refinement, the quantized ZF scheme achieves a comparable performance to the non-linear quantized precoding schemes, where an SNR loss of only 1.5dB is observed compared to the best BER performance achieved by `Constructive, R'.

\begin{figure}[!t]
\centering
\includegraphics[scale=0.45]{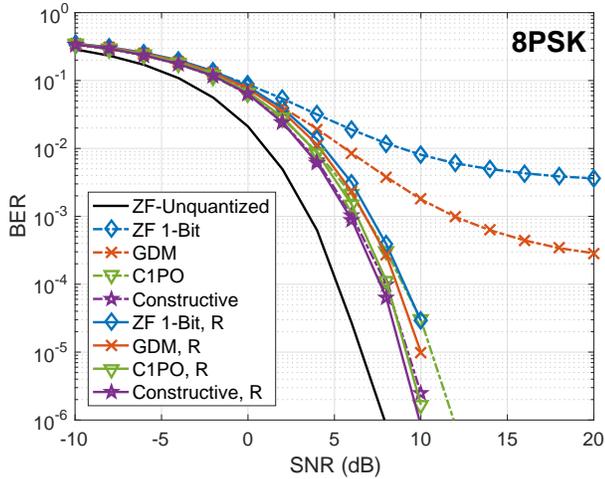}
\caption{BER v.s. transmit SNR, $N_t=128$, $K=8$, 8PSK}
\end{figure}

In Fig. 5, the BER performance with 8PSK modulation for a massive MIMO system with $N_t=128$ transmit antennas and $K=8$ users is depicted. A similar trend to Fig. 3 and Fig. 4 is observed, where the quantized linear ZF method without refinement achieves the worst BER performance, while the refinement process offers the best performance improvement for the quantized ZF scheme. Specifically, there is an SNR loss of only 1dB for `ZF 1-Bit, R' compared to the approach `Constructive, R' that returns the best BER performance. 

Considering the complexity, the quantized linear ZF method with the refinement process achieves the best performance-complexity tradeoff, and is therefore the most promising technique in a practical 1-bit massive MIMO system.

\section{Conclusion}
In this paper, a low-complexity refinement method is proposed for the massive MIMO downlink with 1-bit DACs, which is applicable to any existing quantized precoding methods with PSK modulations. By first formulating the symbol-scaling performance metric based on constructive interference, the refinement process modifies the transmit signal on one antenna at a time and further improves this performance metric with a low computational cost. Numerical results have shown that the refinement offers additional performance improvements for existing schemes, especially for the low-complexity quantized linear ZF method.

\section*{Acknowledgment}
This work was supported by the Royal Academy of Engineering, U.K., the Engineering and Physical Sciences Research Council (EPSRC) project EP/M014150/1, the China Scholarship Council (CSC), and the U.S. National Science Foundation under grant CCF-1703635.

\bibliographystyle{IEEEtran}
\bibliography{refs.bib}

\end{document}